\documentclass[reprint,superscriptaddress,nofootinbib,amsmath,amssymb,aps,pra]{revtex4-1}

\usepackage{color}
\usepackage{braket}
\usepackage{graphicx}
\usepackage{dcolumn}
\usepackage{bm}
\usepackage{bbm}
\usepackage[mathlines]{lineno}

\begin{document}

\title{Advances in sequential measurement and control of open quantum systems}

\author{Stefano Gherardini}
\email{gherardini@lens.unifi.it}
\affiliation{\mbox{Department of Physics and Astronomy, University of Florence,} via G. Sansone 1, I-50019 Sesto Fiorentino, Italy.}
\affiliation{\mbox{LENS and QSTAR,} via N. Carrara 1, I-50019 Sesto Fiorentino, Italy.}

\author{Andrea Smirne}
\email{andrea.smirne@uni-ulm.de}
\affiliation{\mbox{Institute of Theoretical Physics, and IQST, Universit\"at Ulm,} Albert-Einstein-Allee 11, 89069 Ulm, Germany.}

\author{Matthias M. M\"uller}
\email{ma.mueller@fz-juelich.de}
\affiliation{\mbox{LENS and QSTAR,} via N. Carrara 1, I-50019 Sesto Fiorentino, Italy.}
\affiliation{\mbox{Institute for Quantum Control,} Peter Gr\"unberg Institut, Forschungszentrum J\"ulich, 52425 J\"ulich, Germany}

\author{Filippo Caruso}
\email{filippo.caruso@unifi.it}
\affiliation{\mbox{Department of Physics and Astronomy, University of Florence,} via G. Sansone 1, I-50019 Sesto Fiorentino, Italy.}
\affiliation{\mbox{LENS and QSTAR,} via N. Carrara 1, I-50019 Sesto Fiorentino, Italy.}


\begin{abstract}
Novel concepts, perspectives and challenges in measuring and controlling an open quantum system via sequential schemes are shown. We discuss how similar protocols, relying both on repeated quantum measurements and dynamical decoupling control pulses, can allow to: (i) Confine and protect quantum dynamics from decoherence in accordance with the Zeno physics. (ii) Analytically predict the probability that a quantum system is transferred into a target quantum state by means of stochastic sequential measurements. (iii) Optimally reconstruct the spectral density of environmental noise sources by orthogonalizing in the frequency domain the filter functions driving the designed quantum-sensor. The achievement of these tasks will enhance our capability to observe and manipulate open quantum systems, thus bringing advances to quantum science and~technologies.
\begin{description}
\item[Keywords]open quantum systems, quantum Zeno physics, noise quantum sensing.
\end{description}
\end{abstract}

\maketitle

\section{Introduction}

Let us consider a non-isolated quantum mechanical system $\mathbb{S}$, defined within the finite-dimensional Hilbert space $\mathcal{H}_{\mathbb{S}}$. Its dynamics is governed by a time-dependent Hamiltonian of the form $H(t) = H_0 + H_{\rm control}(t)$, where $H_0$ is the Hamiltonian of the system but without taking into account any interaction with the external environment $\mathbb{E}$. Instead, $H_{\rm control}(t) = \lambda(t)\sigma_{\mathbb{S}}$ denotes a coherent control term where $\sigma_{\mathbb{S}}$ is an arbitrary operator acting on $\mathbb{S}$. $H_{\rm control}(t)$ depends on the function $\lambda(t)$ that is properly modulated so as to fulfill the control tasks required by the user. In case $\mathbb{S}$ is a closed system, the Hamiltonian $H_0$ is time-independent, and the only ways to interact with the system are given by performing control actions and measurements, usually on a portion of the system wave-function modeling the system quantum state. Conversely, in case $\mathbb{S}$ is in contact with other external systems\,\cite{PetruccioneBook}, it has to be considered as an open quantum system. The effects of such interactions affect only $\mathbb{S}$ and they can be easily modeled by adding in the Hamiltonian $H_0$ a non-deterministic term proportional to a stochastic field $E(t)$, which can be effectively seen as an environmental noise contribution. Under this hypothesis, the evolution of the system is described by a stochastic quantum dynamics; in this regard, results stemming from the statistical field theory\,\cite{ParisiBook} have been recently derived\,\cite{Gherardini2016NJP,GherardiniQST2017,MuellerANNALEN2017}. However, the noise can involve not only $\mathbb{S}$ but also the control pulse and measurement apparata. Though the effects of such noise sources lead to systematic errors that can be selectively identified and attenuated, they need to be properly modeled so as to avoid a substantial loss of efficiency and accuracy. Otherwise, one can adopt the standard description of open quantum systems, whereby the system is physically coupled to a structured non-equilibrium environment modeling its surroundings. In this case, the global dynamics is governed by an Hamiltonian $H(t)$ including also the term $H_{\rm int}$ that describes the interaction between $\mathbb{S}$ and $\mathbb{E}$. In case $H_{\rm int}$ is fully known and described by deterministic coupling terms, the dynamics of $\mathbb{S}$ is deterministic as well; conversely, by including in $H_{\rm int}$ also the action of stochastic fields, one can recover the stochastic dynamics like that in\,\cite{MuellerANNALEN2017,MuellerSCIREP2016}.

{\bf Repeated quantum measurements}. Let us assume to monitor the dynamics of $\mathbb{S}$ within the time interval $[0,t_{\rm fin}]$, which is defined by $m$ distinct instants $t_{\rm fin} = t_{m} > t_{m-1} > \ldots t_1 > t_0 = 0$, not~necessarily equidistributed in time. Protocols allowing for this purpose are given by a sequence of quantum measurements, locally performed on $\mathbb{S}$ and in correspondence of
$t_k$, $k = 1,\ldots,m$, according to the observables $\mathcal{O}_k \equiv \mathcal{O}_{\mathbb{S},k} \otimes I_{\mathbb{E}}$, where $\mathcal{O}_{\mathbb{S},k} \equiv F_{\theta_k}$ and $I$ denotes the identity operator. Specifically, $\{\theta_k\}$ is the set of the possible measurement outcomes, while $\{F_{\theta}\}$ denotes the set of positive Hermitian semi-definite operators on $\mathcal{H}_{\mathbb{S}}$ satisfying the relation $\sum_{\theta}F_{\theta} = I_{\mathbb{S}}$. Given the system density matrix $\rho_{\mathbb{S},k}$ at time $t_k$, the probability that the outcome $\theta$ associated with the measurement operator $F_{\theta}$ occurs is returned by the trace ${\rm Tr}[\rho_{\mathbb{S},k}F_{\theta}]$, while the post-measurement state of $\mathbb{S}$ equals to
\mbox{$\widetilde{\rho}_{\mathbb{S},k} = (M_{\theta}\rho_{\mathbb{S},k}M_{\theta}^{\dagger})/{\rm Tr}[M_{\theta}\rho_{\mathbb{S},k}M_{\theta}^{\dagger}]$}, where $M_{\theta}$ fulfills the identity $F_{\theta} = M^{\dagger}_{\theta}M_{\theta}$ (notice that for the same value of $\theta$ two different operators $F_{\theta}$ are not allowed).

{\bf Coherent pulsed control couplings}. Coherent (open-loop) control pulses are an essential tool to efficiently perform quantum sensing\,\cite{Giovannetti_NatPhon_2011,Degen_RevModPhys_2017}. Here, we briefly introduce the control strategy also called \textit{Dynamical Decoupling} (DD)\,\cite{Viola_PRL_1999}, which is given by applying a sequence of $\pi-$pulses, i.e., short and strong control pulses that invert the phase of the quantum system $\mathbb{S}$ - usually a qubit - used as a sensor. In the noise sensing context, the qubit-sensor is placed in interaction with an external stochastic field $E(t)$ with the aim to infer its fluctuation profile. For this purpose, the qubit is prepared in the ground state $|0\rangle$ and a $\pi/2-$pulse is firstly performed so as to transfer the system in the superposition state $(|0\rangle+|1\rangle)/\sqrt{2}$, where $|1\rangle$ is the corresponding excited state. Only at this point the sequence of $\pi-$pulse is applied to $\mathbb{S}$, which thus acquires a phase $\phi(t)$, providing us (if measured) information about the fluctuating field. In fact, at the end of the DD sequence at time $t_{\rm fin}$, the state of the qubit-sensor is $[\mathrm{e}^{i\phi(t_{\rm fin})}|0\rangle + \mathrm{e}^{-i\phi(t_{\rm fin})}|1\rangle]/\sqrt{2}$, where
$\phi(t_{\rm fin}) \equiv \int_0^{t_{\rm fin}} y(t') E(t') dt'$ and $y(t)\in\{-1,1\}$ is the control pulse modulation function that switches sign whenever a $\pi-$pulse occurs. Finally, a second $\pi/2$-pulse brings the qubit-sensor into the final state $[(\mathrm{e}^{i\phi(t_{\rm fin})}+\mathrm{e}^{-i\phi(t_{\rm fin})})|0\rangle + (\mathrm{e}^{i\phi(t_{\rm fin})}-\mathrm{e}^{-i\phi(t_{\rm fin})})|1\rangle]/2$ and the probability $p_{|0\rangle}(t_{\rm fin})$ that the qubit-sensor is in the state $|0\rangle$ at time $t_{\rm fin}$ is measured.

\section{Stochastic Quantum Zeno physics}

The main purpose to apply sequences of quantum measurements based on the quantum Zeno physics\,\cite{Misra1977,ItanoPRA1990,FisherPRL2001,FacchiPRL2002,FacchiJPA2008,SmerziPRL2012} is to force the dynamics of $\mathbb{S}$ to be confined within the Hilbert subspace defined by the measurement observable. Since their introduction, standard observation protocols, given by sequences of repeated projective measurements, have been applied to closed quantum systems by assuming that between each measurement the system freely evolves with unitary dynamics for a constant small time interval $\tau$. More formally, all the measurement observables $\mathcal{O}_{\mathbb{S},k}$ are set equal for any $k$ to the projector (Hermitian, idempotent operator, in general with dimension greater than 1) $\Pi$ that defines the confinement Hilbert subspace $\mathcal{H}_{\mathbb{S}}^{\Pi} \equiv \Pi\mathcal{H}_{\mathbb{S}}$. In $\mathcal{H}_{\mathbb{S}}^{\Pi}$ the dynamics of the system is described exclusively by the projected (or Zeno) Hamiltonian $\Pi H\Pi$, as it has been observed in Refs.\,\cite{SchaferZeno,Signoles2014}.

Recently, the Probability Density Functions (PDF) $p(\tau)$ and $p(E(t))$ respectively of the time intervals $\tau$ between measurements or of the stochastic field $E(t)$ have been taken into account and the acronym \textit{Stochastic Quantum Zeno Dynamics} (SQZD) has been introduced\,\cite{MuellerANNALEN2017,MuellerSCIREP2016}. Here, the peculiarity is that the interaction model with $\mathbb{E}$ is given by fluctuation profiles of one or more parameters entering in the dynamics of $\mathbb{S}$. Such a noise involves a no longer effective confinement of the system dynamics, but~on the other side it increases the capability of the system to explore a larger number of configurations within $\mathcal{H}_{\mathbb{S}}^{\Pi}$. In a controlled setup this viewpoint makes emerge the noise, \mbox{i.e.,\,the presence} of an external environment, as a \textit{resource}\,\cite{GherardiniPhDThesis,GherardiniPRE2018}. In this regard, a first result is given by the analytical expression of the probability distribution that $\mathbb{S}$ belongs to the confinement subspace $\mathcal{H}_{\mathbb{S}}^{\Pi}$ after a large enough number of sequential measurements. In particular, let us define the survival probability $\mathcal{P}_{t_{\rm fin}} \equiv {\rm Prob}(\rho_{\mathbb{S},t_{\rm fin}}\in\mathcal{H}_{\mathbb{S}}^{\Pi})$ to find the system in the confinement subspace. In case the time intervals $\tau_j$ between measurements are independent and identically distributed random variables, $\mathcal{P}_{t_{\rm fin}}$ is equal to
\begin{equation}
\mathcal{P}_{t_{\rm fin}} = \prod_{j=1}^{m}{\rm Tr}\left[\Pi U_{j-1:j}\Pi\rho_{\mathbb{S},t_{j}}\Pi U_{j-1:j}^{\dagger}\right],
\end{equation}
where $U_{j-1:j} \equiv \hat{T} \exp\left(-(i/\hbar)\int_{t_{j-1}}^{t_{j}}H(t)dt\right)$, $\hat{T}$ is the time ordering operator, while $m$ denotes the total number of projective measurements applied to $\mathbb{S}$. Being able to take values from an ensemble of configurations, the density matrix of $\mathbb{S}$ at time $t_{\rm fin}$ and the survival probability $\mathcal{P}_{t_{\rm fin}}$ are random quantities. Thus, the prediction power of the method is constrained by our ability to compute the most probable value $\mathcal{P}^{\ast}$ of the survival probability after a single realization of the sequence of measurements. In this regard, by using the Large Deviation (LD) theory, it has been derived also the analytical expression of $\mathcal{P}^{\ast}$ for a large enough value of $m$ \cite{MuellerANNALEN2017}, i.e.,
\begin{equation}\label{P_star}
\mathcal{P}^{\ast} = \exp\left(m\int_{\tau,\eta(t)}p(\tau)\,p(\eta(t))\ln(q(\tau,\eta(t)))\,d\tau\,d\eta(t)\right),
\end{equation}
where $q(\tau,\eta(t))$ is a functional identifying the probability that $\mathbb{S}$ belongs to $\mathcal{H}_{\mathbb{S}}^{\Pi}$ at time $t$ after the application of a couple of projective measurements interspersed by the time interval $\tau$. In~the limit of small $\tau$'s, $q(\tau,\eta(t))$ admits the second-order expansion $q(\tau,\eta(t)) \approx 1 - \eta(t)^2\tau^2$. In~particular, $\eta(t) \equiv \Delta_{\rho_{\mathbb{S},t}^{\Pi}}H_{\Pi}(t)$ is the standard deviation of $H_{\Pi}(t) \equiv H(t) - \Pi H(t)\Pi$ with respect to the system density matrix within the confinement subspace, i.e.,\,$\rho_{\mathbb{S},t}^{\Pi} \equiv U_{t-\tau:t}^{\Pi}\rho_{\mathbb{S},t-\tau}^{\Pi}(U_{t-\tau:t}^{\Pi})^{\dagger}$ with $U_{t-\tau:t}^{\Pi} \equiv \hat{T} \exp\left(-(i/\hbar)\int_{t-\tau}^{t}\Pi H(t')\Pi dt'\right)$. Moreover, in Eq.\,(\ref{P_star}) $p(\eta(t))$ denotes an artificial PDF obeying the relation
$\int p(\eta(t))\eta(t)^{2}d\eta(t) = \frac{1}{t_{\rm fin}}\int_{0}^{t_{\rm fin}}\Delta^{2}_{\rho_{\mathbb{S},t}^{\Pi}}H_{\Pi}(t)dt$, that fixes on average the leakage dynamics of $\mathbb{S}$ outside the confinement subspace $\mathcal{H}_{\mathbb{S}}^{\Pi}$.

SQZD is a special class of dynamics induced by protocols based on sequential measurements. In~the more general case, the measurement observables within the sequence are no longer equal to a single projector, and also the presence of coherent control terms in the Hamiltonian $H(t)$ can be taken into account. Such protocols are expected to provide the proper tool to explore the whole Hilbert space of a quantum system by engineering the occurrence of the measurement operators in specific time instants, so as to move the system population from one portion of the Hilbert space to another. This question is still challenging, because it requires to properly modulate a control pulse $\lambda(t)$ so that the probability distribution of $\mathcal{P}_{t_{\rm fin}}$ is peaked in correspondence of a target value chosen by the user. As further remark, let us also observe that coherent dynamical couplings with an auxiliary system have been studied as tools playing the role of a measurement. In other words the back-action induced by a quantum measurement has been reproduced by using a coherent pulse, and an equivalence between sequences of repeated measurements and pulsed control couplings have been established. However, although such an approximation revealed to experimentally work quite well in peculiar dynamical conditions, as shown for example in\,\cite{SchaferZeno}, one has always to keep in mind that, even when the measurement outcomes are not recorded, sequential measurements can lead to dissipative dynamics, thus implying loss of quantum coherence. On the other side, pulsed control couplings are not always able to reproduce the measurement back-action; e.g.,\,the ideal confinement of quantum dynamics in ensured only by sequence of projective measurements\,\cite{Gherardini2016NJP,MuellerANNALEN2017}.

\section{Noise-Robust Quantum Sensing}

The prediction power of the results shown in the previous section is ensured by knowing (also~partially) the fluctuation profiles of the parameters entering in the dynamics of $\mathbb{S}$. In this regard, noise sensing (or spectroscopy) aims to determine the spectral density of the noise originated by the interaction between the quantum system $\mathbb{S}$ used as a probe and its external environment $\mathbb{E}$\,\cite{Degen_RevModPhys_2017}.

Let us simply take a qubit as sensing device to detect the presence of stochastic (time-varying) magnetic fields $E(t)$. Thus, assuming that the qubit-sensor is coherently manipulated, the application of different and optimized sequences of control pulses allows to enhance the sensor sensitivity in probing the target field\,\cite{Kofman_PRL_2001,Paz-SilvaPRL2014,PoggialiPRX2018,Mueller_sensing_2018}. In this regard, the DD control strategy has been successfully applied to noise sensing\,\cite{Alvarez_PRL_2011}. Its main peculiarity is that the decay rate (or decoherence function) $\chi(t)$ of the qubit-sensor due to the presence of $\mathbb{E}$ is related to the probability $p_{|0\rangle}(t_{\rm fin})$ via the equation $\chi(t) = -\ln(1-2p_{|0\rangle}(t_{\rm fin}))$ (see e.g.,\,\cite{Mueller_sensing_2018}), and is simply given by the overlap in the frequency domain between the environmental spectral density function $S(\omega)$ and the filter function $F(\omega)$ of the DD sequence driving the qubit. More~specifically
\begin{equation}
\chi(t) = \frac{1}{2}\int_0^t\int_0^t y(t')y(t'')g(t'-t'')dt'dt'',
\end{equation}
where $y(t)$ is the control modulation function and
\begin{eqnarray}
g(t'-t'') &\equiv&\int_{-\infty}^{\infty}\int_{-\infty}^{\infty}p(E(0),E(t'-t''))\nonumber \\
&\times& E(0)E(t'-t'')dE(0)dE(t'-t'')
\end{eqnarray}
is the autocorrelation function $\big\langle E(0)E(t'-t'')\big\rangle$ of the fluctuating field $E(t)$, with $p(E(0),E(t'-t''))$ denoting the joint PDF of $E(t)$ in the two time instants $t=0$ and $t=t'-t''$. The previous equation is valid if we assume that the mean value of $E(t)$ is equal to zero, i.e., $\big\langle E(t)\big\rangle = 0$, and $E(t)$ is a stationary process so that $g(t',t'') = g(t'-t'')$. Then, being $S(\omega)$ and $Y(\omega)$ defined respectively as the Fourier transform of the autocorrelation function $g(t)$ and the pulse modulation function $y(t)$, one has that
\begin{equation}
\chi(t) = \int_{-\infty}^{\infty}S(\omega)F(\omega)d\omega,
\end{equation}
where $F(\omega) \equiv \frac{4}{\pi}|Y(\omega)|^2$ ($F(\omega)$ is usually called filter function). As a result, from the measurement of $p_{|0\rangle}$ at the end of the protocol, one can obtain the corresponding value of the decoherence function at time $t_{\rm fin}$, i.e., $\chi(t_{\rm fin})$.

As second step, different filter functions $F(\omega)$ can be designed by engineering the pulse modulation function $y(t)$, with the aim to reconstruct $S(\omega)$ in a range $\omega\in[0,\omega_c]$ for a given cut-off frequency $\omega_c$. To this end, let us consider a set of $N$ filter functions $F_{n}(\omega)$, $n = 1,\dots,N$, generated by equidistant $\pi-$pulse sequences with a different number of pulses placed in correspondence of the zeros of $\cos[\omega_{max}\frac{n-1}{N}t']$. As given by the \textit{Filter Orthogonalization} (FO) protocol, introduced in \cite{Mueller_sensing_2018} for noise-robust quantum sensing, we quantify the overlap between the $N$ filter functions $F_{n}(\omega)$ in the frequency domain, by computing the following $N\times N$ symmetric matrix $A$ with matrix elements
\begin{equation}
A_{nl} \equiv \int_0^{\omega_c} F_n(\omega)F_l(\omega)\,d\omega.
\end{equation}
An accurate estimate of $S(\omega)$ is then obtained in case of no overlaps between the $F_n(\omega)$'s, i.e., if the filter functions are orthogonal and they span a $N-$dimensional space. Otherwise, we orthogonalize the matrix $A$ by using the transformation $V A V^{\dagger} = \mathrm{diag}(\lambda_1,\dots,\lambda_N)$, where $V$ is an orthogonal matrix and $\lambda_n$ are the eigenvalues of $A$. In this way, we will determine a transformed version of the filter functions $F_n(\omega)$, i.e.,
\begin{equation}
\widetilde{F}_{n}(\omega) = \frac{1}{\sqrt{\lambda_n}}\sum_{l=1}^N V_{nl}F_l(\omega),\quad n = 1,\dots,N,
\end{equation}
that are all orthogonal functions for any (integer) value of $n$. The procedure is concluded by expanding $S(\omega)$ in the transformed orthogonal basis, so that also the $\chi(t_{\rm fin})$'s are accordingly modified in the transformed coefficients
\begin{equation}
\widetilde{\chi}_{n} \equiv \int_0^\infty S(\omega) \widetilde{F}_n(\omega)=\frac{1}{\sqrt{\lambda_n}}\sum_{l=1}^N \chi_{l}(t_{\rm fin})V_{nl},
\end{equation}
and the estimate of $S(\omega)$, i.e., $\widetilde{S}(\omega)$, is simply given by the following expansion:
\begin{eqnarray}
\widetilde{S}(\omega) = \sum_{n=1}^{N}\widetilde{\chi}_{n}\widetilde{F}_{n}\,.
\end{eqnarray}
To sum-up, the FO protocol is a reconstruction algorithm for the spectrum of a signal that is based on the orthogonalization of the applied filter functions, each of them corresponding to a properly engineered pulse modulation function -- usually $\pi-$pulses. The filter functions select specific frequency ranges of the power spectral density $S(\omega)$, and, in order to correctly estimate the functional behaviour of $S(\omega)$, a set of $K$ \emph{orthogonal} filter functions $F_{k}(\omega)$ has to be employed. However, due to physical and/or experimental constraints, a sufficiently large number of orthogonal filter functions cannot be realized, and, mainly for this reason, the FO protocol aims to solve this issue and thus speed-up the sensing procedure of different forms of noise.

\section{Conclusions and Outlook}

In conclusion, we have shown two novel approaches of measurement and control theory, respectively based on the physics of Zeno phenomena\,\cite{Gherardini2016NJP,GherardiniQST2017,MuellerANNALEN2017} and noise-sensing spectroscopy\,\cite{Mueller_sensing_2018}. The first relies on applying repeated quantum measurements, while the second on performing DD sequences and using models with fluctuating parameters entering in the dynamics of $\mathbb{S}$. Such methods are believed to be a concrete step towards the realization of novel quantum technologies, especially quantum-based sensing devices for future biomedicine applications.

As main outlook, it is worth analyzing with the same formalism the \textit{non-Markovianity} (NM)\,\cite{Rivas_Review_2014,Breuer_Review_2016} of the open quantum system $\mathbb{S}$ in reference to the multi-time statistics obtained by locally measuring $\mathbb{S}$\,\cite{Smirne2017Coherence,Gherardini_Non-Markov}. Indeed, in case our knowledge of the system-environment interaction is a-priori unknown (or partially known), our capability to evaluate the NM of the system dynamics is simply given by the outcomes from a sequence of measurements, i.e., by monitoring the change of the state of the system due to the presence of $\mathbb{E}$. In the quantum mechanical context, this is still a challenging issue, since it requires to understand which is the role and the effects on the dynamics of the measurement back-action in probing the NM of $\mathbb{S}$.

\textbf{Acknowledgments}:
The authors thank Susana Huelga for useful discussions and for the reading of the manuscript. This work was financially supported from the Fondazione CR Firenze through the project Q-BIOSCAN.

\end{document}